\documentclass{JHEP3}

\usepackage{mathrsfs}
\usepackage{amsmath}
\usepackage{amsfonts}
\usepackage{cite}
\usepackage{subfigure,epsfig}

\title{MadDipole:
Automation of the Dipole Subtraction Method in MadGraph/MadEvent}

\author{R.\ Frederix\\
Center for Particle Physics and Phenomenology (CP3),\\
Universit\'{e} Catholique de Louvain, \\
Chemin du Cyclotron 2, 1348 Louvain-la-Neuve, Belgium\\
}
\author{T.\ Gehrmann, N.\ Greiner\\
Institut f\"ur Theoretische Physik,\\
Universit\"at Z\"urich,\\
Winterthurerstrasse 190, 8057 Z\"urich, Switzerland\\
}
\preprint{CP3-08-39\\ZU-TH 13/08}
\abstract{
We present the implementation of the dipole subtraction formalism for
the real radiation contributions to any next-to-leading order QCD process
in the MadGraph/MadEvent
framework. Both massless and massive dipoles are considered. Starting
from a specific $(n+1)$-particle process the package provides a
Fortran code for all possible dipoles to all Born processes that
constitute the subtraction term to the $(n+1)$-particle process.  The
output files are given in the usual ``MadGraph StandAlone'' style
using helicity amplitudes.
}

\begin{document}
\newcommand{\bra}[1]{\langle#1|}
\newcommand{\ket}[1]{|#1\rangle}
\newcommand{\nn}{\nonumber}
\section{Introduction}

Physics studies at the upcoming CERN LHC collider will frequently
involve multi-particle final states. Especially searches for physics
beyond the standard model rely on the reconstruction of new particles
from their decay products, often through decay chains. Equally,
requiring accompanying particles in the final state may serve to
improve the ratio of signal to background processes, as done for
example in the Higgs boson search through the vector boson fusion
channel.  Meaningful searches for these signals require not only a
very good anticipation of the expected signal, but also of all
standard model backgrounds yielding identical final state
signatures. From the theoretical point of view, high precision implies
that one has to go beyond the leading order in perturbation theory to
be able to keep up with the precision of the measurements.\\
For leading order processes there have been many developments
concerning event generation and simulation tools in the last two
decades such as MadGraph/MadEvent
\cite{Stelzer:1994ta,Maltoni:2002qb,Alwall:2007st} CompHEP/CalcHEP
\cite{Boos:2004kh}/\cite{Pukhov:2004ca}, SHERPA
\cite{Gleisberg:2003xi} and WHIZARD \cite{Kilian:2001qz} and also
programs using different approaches such as ALPGEN
\cite{Mangano:2002ea} and HELAC \cite{Kanaki:2000ey}.  All these
programs are multi-purpose event generator tools, which are able to
compute any process (up to technical restrictions in the multiplicity)
within the standard model, or within alternative theories specified by
their interaction Lagrangian or Feynman rules. They usually provide
event information which can be interfaced into parton shower,
hadronization and/or detector simulation. \\
Next-to-leading order (NLO) calculations are at present performed on a
process-by-process basis. The widely-used programs MCFM
\cite{Campbell:1999ah,Campbell:2002tg}, NLOJET++ \cite{Nagy:2003tz},
MC@NLO \cite{Frixione:2002ik,Frixione:2006gn} and programs based on
the POWHEG method
\cite{Nason:2004rx,Nason:2006hfa,LatundeDada:2006gx,Frixione:2007nw,Alioli:2008gx,Hamilton:2008pd}
collect a variety of different processes in a standardized framework,
the latter two methods also match the NLO calculation onto a parton
shower. \\
The NLO QCD corrections to a given process with a $n$-parton final 
state receive two types of contributions: the one-loop virtual 
correction to the $(2\to n)$-parton scattering process, and the real emission 
correction from all possible $(2\to n+1)$-parton scattering processes. 
For the numerical evaluation,  one has to be able to 
compute both types of contributions separately. \\ 
The computation of one-loop corrections to multi-particle scattering 
amplitudes was performed on a case-by-case basis up-to-now, the 
calculational complexity increased considerably with increasing number 
of external partons. Since only a limited number of one-loop integrals 
can appear~\cite{Melrose:1965kb,Bern:1993kr} in the final result, 
the calculation of one-loop corrections 
can be reformulated as determination of the coefficients of these 
basis integrals, plus potential rational terms. 
Enormous progress~\cite{Bern:1994zx,Bern:1995db,Bern:2005cq,Berger:2006ci,Forde:2007mi,Denner:2005fg,Denner:2005nn,Binoth:2005ff,Binoth:2006hk,Britto:2006sj,Mastrolia:2006ki,Britto:2007tt,Britto:2008vq,Anastasiou:2006gt,Ellis:2007br,Giele:2008ve,Ellis:2008ir,Ossola:2006us,Ossola:2008xq,Mastrolia:2008jb}
 has been made in the recent past in 
the systematic determination of the one-loop integral coefficients and 
rational terms, and  steps towards fully automated programs 
for the calculation of one-loop multi-parton amplitudes were made 
with the packages CutTools~\cite{Ossola:2007ax}, 
BlackHat~\cite{Berger:2008sj}, Rocket~\cite{Giele:2008bc}
 and GOLEM~\cite{Binoth:2008gx}.\\
The real emission corrections contain soft and collinear singularities, which
become explicit only after integration over the appropriate real radiation 
phase space yielding a hard $n$-parton final state. They are canceled 
by the singularities from the virtual one-loop contributions, thus 
yielding a finite NLO correction. 
To systematically extract the real radiation singularities from arbitrary 
processes, a variety of methods, based either on phase-space 
slicing~\cite{Giele:1991vf}
or on the introduction of process-independent subtraction 
terms~\cite{Kunszt:1992tn} have been proposed. Several 
different algorithms to derive subtraction terms  are available: 
 residue subtraction~\cite{Frixione:1995ms},
dipole subtraction~\cite{Catani:1996vz,Catani:2002hc} and  antenna 
subtraction~\cite{Kosower:1997zr,Campbell:1998nn,GehrmannDeRidder:2005cm,Daleo:2006xa}. \\
Especially the dipole subtraction formalism, which provides local subtraction 
terms for all possible initial and final state 
configurations~\cite{Catani:1996vz} and allows to account for radiation 
off massive partons~\cite{Catani:2002hc}, is used very widely in NLO 
calculations. 
It has recently also been 
automated in the SHERPA framework \cite{Gleisberg:2007md} and
the TeVJet framework \cite{Seymour:2008mu}, and most recently 
in the form of independent libraries~\cite{Hasegawa:2008ae} interfaced to 
MadGraph. The dipole subtraction within the 
SHERPA framework is  
not available as a stand-alone tool, while  within the TeVJet framework, 
the user
needs to provide all the necessary process dependent information. Moreover
both approaches have only included massless particles for the dipoles.
There is still no general tool available which is able to produce
the dipole terms for an arbitrary process and which can also deal with
massive partons.\\
In this paper we present MadDipole, an automatic generation of the dipole 
subtraction terms using
the MadGraph/MadEvent framework. The results are {\tt Fortran}
 subroutines which
return the squared amplitude for all possible dipole configurations in
the usual MadGraph style. We describe the construction of the 
dipole subtraction terms and their implementation
in Section~\ref{sec:dipole}. Results from various checks of the 
implementation are provided Section~\ref{sec:checks}, and
instructions on the practical usage of the package are  contained in 
Section~\ref{sec:use}.

\section{Construction of dipole terms}
\label{sec:dipole}
The fundamental building blocks of the subtraction terms in the dipole 
formalism~\cite{Catani:1996vz,Catani:2002hc} are dipole 
splitting functions ${\bf V}_{ij,k}$, which
involve only three partons: emitter $i$, unresolved parton $j$, 
spectator $k$. A dipole splitting function accounts for the collinear limit of 
$j$ with $i$, and for part of the soft limit of $j$ in between $i$ and $k$. 
The dipole factors, which constitute the subtraction terms, 
are obtained by multiplication with 
 reduced matrix elements, where partons $i$, $j$
and $k$ are replaced by recombined 
pseudo-partons $\widetilde{ij}$, $\widetilde{k}$.
The full soft behavior is recovered after summing all dipole factors.  \\
Throughout the whole paper we are using the notation introduced in Refs.~\cite{Catani:1996vz}
and \cite{Catani:2002hc}. Independent on
whether we have initial or final state particles we can write an arbitrary
dipole in the form
\begin{equation}
\label{dipole}
 {\cal{D}}_{ij,k}\quad \sim \quad _m\bra{1,...\tilde{ij},...,\tilde{k},...,m+1}
\frac{{\bf T}_k\cdot {\bf T}_{ij}}{{\bf T}_{ij}^2}{\bf V}_{ij,k}\ket{1,...\tilde{ij},...,\tilde{k},...,m+1}_m.
\end{equation}
The amplitude factors $\bra{\ldots}$ (`bra') and $\ket{\ldots}$ (`ket')
on the right hand side are tensors in color space.
The helicities of the external particles in them are a priori fixed
(but can be summed over for unpolarized processes), while 
the helicities of the pseudo-partons have to be summed over 
after contraction with the dipole splitting function. \\
These Born-level amplitude factors are provided by the usual MadGraph
code. The two elements that combine the ket with the bra are the 
additional color structure $\frac{{\bf T}_k\cdot {\bf T}_{ij}}{{\bf T}_{ij}^2}$ and the
dipole splitting function ${\bf V}_{ij,k}$.

\subsection{Color and helicity management}
For the calculation of the color factors there already exist routines
in the MadGraph program. Our intension was to use exactly these routines because
this code is very well-confirmed and efficient. We have included the additional
color operator, ${\bf T}_k\cdot {\bf T}_{ij}$, by rewriting the internal MadGraph
color labelling for the ket-side only.
After insertion of this color operator the color structure is no longer multiplied by its own complex
conjugate  and therefore the routine that squares the color needed to be altered,
to multiply the modified ket by its original complex conjugate.
We emphasize that due to the factorial growth of the color factors MadGraph
can not handle more than seven colored particles.\\ \\
For the insertion of the splitting function ${\bf V}_{ij,k}$ several changes
with respect to the original code are required.
One has to keep in mind that in general the splitting function is a tensor in
helicity space, {\it i.e.},
\begin{equation*}
 {\bf V}_{ij,k} \equiv \bra{\mu}{\bf V}_{ij,k}\ket{\nu}={\bf V}_{ij,k}^{\mu \nu}.
\end{equation*}
As MadGraph deals with helicity amplitudes, we have to write the dipole in a
slightly different way to be
able to include the calculation of the splitting function in the code.
Neglecting the color for a moment we start from the definition of the dipole
in (\ref{dipole}) and by inserting a full set of 
helicity states $-g_{\mu\nu}=\sum_{\lambda}\epsilon^{*}_{\mu}(\lambda)\epsilon_{\nu}(\lambda)$ we get
\begin{align}
 {\cal{D}}_{ij,k} \ &\sim \ _m\bra{1,...\tilde{ij},...,\tilde{k},...,m+1}_{\mu}
{\bf V}_{ij,k}^{\mu \nu}\  _\nu \ket{1,...\tilde{ij},...,\tilde{k},...,m+1}_m \\
&=\ _m\bra{1,...\tilde{ij},...,\tilde{k},...,m+1}_{\mu'}\left(-g^{\mu'}_{\ \mu}\right)
{\bf V}_{ij,k}^{\mu \nu}\left(-g^{\ \nu'}_{\nu}\right) {}_{\nu'}
\ket{1,...\tilde{ij},...,\tilde{k},...,m+1}_m \notag\\
&=\ \sum \limits_{\lambda_a,\lambda_b}\ _m\bra{\ldots}_{\mu'}\ \epsilon^{*\mu'}(\lambda_b)
\epsilon_{\mu}(\lambda_b) {\bf V}_{ij,k}^{\mu \nu}\ \epsilon^{*}_{\nu}(\lambda_a)\epsilon^{\nu'}(\lambda_a)\ 
_{\nu'}\ket{\ldots}_m  \notag\\
&=\ \sum \limits_{\lambda_a,\lambda_b}\ _m\bra{\ldots}_{\lambda_b}\
V(\lambda_b,\lambda_a)\ _{\lambda_a}\ket{\ldots}_m\notag
\end{align}
with $ V(\lambda_b,\lambda_a)=\epsilon_{\mu}(\lambda_b) \ {\bf V}_{ij,k}^{\mu \nu}\ \epsilon^{*}_{\nu}(\lambda_a)$
and $\epsilon^{\mu}(\lambda)\,{}_{\mu}\ket{\ldots}_m={}_{\lambda}\ket{\ldots}_m$.\\
The polarization vectors are calculated using the HELAS routines \cite{Murayama:1992gi}
already available in the MadGraph code.
The parts that are diagonal in helicity space are trivial to calculate in that sense
that one only has to multiply the MadGraph output for the squared amplitude for a 
given helicity combination with the splitting function. To calculate the off-diagonal
helicity terms, the amplitude for each helicity combination is stored and then
combined with the according amplitude with opposite helicity.\\
For the calculation of the splitting functions and for the remaping of
the momenta we use modified versions of the routines used in 
MCFM~\cite{Campbell:1999ah,Campbell:2002tg}.

\subsection{Massive particles}
\label{sec:masses}
If some of the masses of the external particles are non-zero, in
particular for processes involving top and/or bottom quarks, there are
dipoles for which the unresolved parton is massive. In this case the
collinear singularities are regulated by the mass of the
unresolved parton and the unsubtracted matrix element does therefore
no longer diverge in these collinear limits, but only develops
potentially large logarithms.  Our code still calculates all possible
dipoles, also in which the unresolved parton is massive, but puts them
in a separate subroutine, \texttt{dipolsumfinite(...)}, that is not
evaluated by default. In the limit of large center of momentum energy
or, similarly, small external masses, the user can easily include the
non-divergent dipoles to subtract the associated large logarithms,
which can then be included analytically through the integrated
subtraction terms.  In the limit of zero external masses we have
checked that the results obtained after summing all dipoles are the
same as obtained by generating the code with massless particles from
the start.

\subsection{Phase space restriction}\label{PSrestriction}
The calculation of the subtraction terms is only necessary in the
vicinity of a soft and/or collinear limit. Away from these limits the
amplitude is finite and there is in principle no need to calculate the
computationally heavy subtraction terms.  The distinction between
regions near to a singularity from regions without need for a
subtraction can be parameterized by a parameter
usually labelled $\alpha$ with $\alpha \in [0,1]$, which was
introduced in Ref.~\cite{Nagy:1998bb} for processes involving partons only
in the final state. The case with incoming hadrons, {\it i.e.}, with
partons in the initial state, is described in Ref.~\cite{Nagy:2003tz}.\\
Using the notation of Ref.~\cite{Nagy:2003tz}, the contribution from
the subtraction term to the differential cross section can be written
as
\begin{eqnarray} \nn
\lefteqn{ d\sigma_{ab}^A = \sum_{\{n+1\}} 
d\Gamma^{(n+1)}(p_a,p_b,p_1,...,p_n+1)
  \frac{1}{S_{\{n+1\}}}}\\ \nn
  &&\quad\times\Bigg\{\sum_{\mathrm{pairs}\atop i,j} \sum_{k\neq i,j}
  {\cal D}_{ij,k}(p_a,p_b,p_1,...,p_{n+1})
  F_J^{(n)}(p_a,p_b,p_1,..,\tilde{p}_{ij},\tilde{p}_{k},..,p_{n+1})
  \Theta(y_{ij,k} < \alpha)\\ \nn
&&\quad\qquad+ \sum_{\mathrm{pairs}\atop i,j}
  \bigg[{\cal D}_{ij}^a(p_a,p_b,p_1,...,p_{n+1})
  F_J^{(n)}(\tilde{p}_a,p_b,p_1,..,\tilde{p}_{ij},..,p_{n+1}) 
  \Theta(1-x_{ij,a} < \alpha) \\ \nn
&& \hspace{2cm}
+ (a\leftrightarrow b)\bigg]\\ \nn
&&\quad\qquad+ \sum_{i\neq k}\left[
  {\cal D}_k^{ai}(p_a,p_b,p_1,...,p_{n+1})
  F_J^{(n)}(\tilde{p}_a,p_b,p_1,..,\tilde{p}_k,..,p_{n+1}) 
  \Theta(u_{i} < \alpha) + (a\leftrightarrow b)\right]\\ 
&&\quad\qquad+ \sum_{i}\left[
  {\cal D}^{ai,b}(p_a,p_b,p_1,...,p_{n+1})
  F_J^{(n)}(\tilde{p}_a,p_b,\tilde{p}_1,...,\tilde{p}_{n+1}) 
  \Theta(\tilde{v}_i < \alpha) + (a\leftrightarrow b)\right]\Bigg\}\;\;.
\nn \\  \label{dipole-terms}
\end{eqnarray}
The functions ${\cal D}_{ij,k}$, ${\cal D}_{ij}^a$, ${\cal D}_k^{ai}$
and ${\cal D}^{ai,b}$ are the dipole terms for the various
combinations for emitter and spectator.  $\sum_{ \{ n+1\}}$ denotes
the summation over all possible configurations for this
$(n+1)$-particle phase space which is labelled as $d\Gamma^{(n+1)}$
and the factor $S_{\{n+1\}}$ is the symmetry factor for identical
particles. We have introduced four different $\alpha$-parameters, one
for each type of dipoles. In our code they are called
\texttt{alpha\_ff}, \texttt{alpha\_fi}, \texttt{alpha\_if} and
\texttt{alpha\_ii} for the final-final, finial-initial, initial-final
and initial-initial dipoles, respectively. The actual values for these
parameters are by default set to unity, corresponding to the original
formulation of the dipole subtraction
method~\cite{Catani:1996vz,Catani:2002hc}, but can be changed by the
user in the file \texttt{dipolsum.f}.\\
It has to be kept in mind that the integrated dipole factors, which
are to be added with the virtual $n$-parton contribution, will also
depend on $\alpha$. For case of massless partons, the
$\alpha$-dependence of the integrated terms is stated in
\cite{Nagy:1998bb,Nagy:2003tz} while for massive partons results for
most cases can be found in \cite{Campbell:2004ch,Campbell:2005bb}.

\section{Checks}
\label{sec:checks}
\begin{figure}[t]
  \begin{center}
    \mbox{
      \subfigure[]{
        \epsfig{file=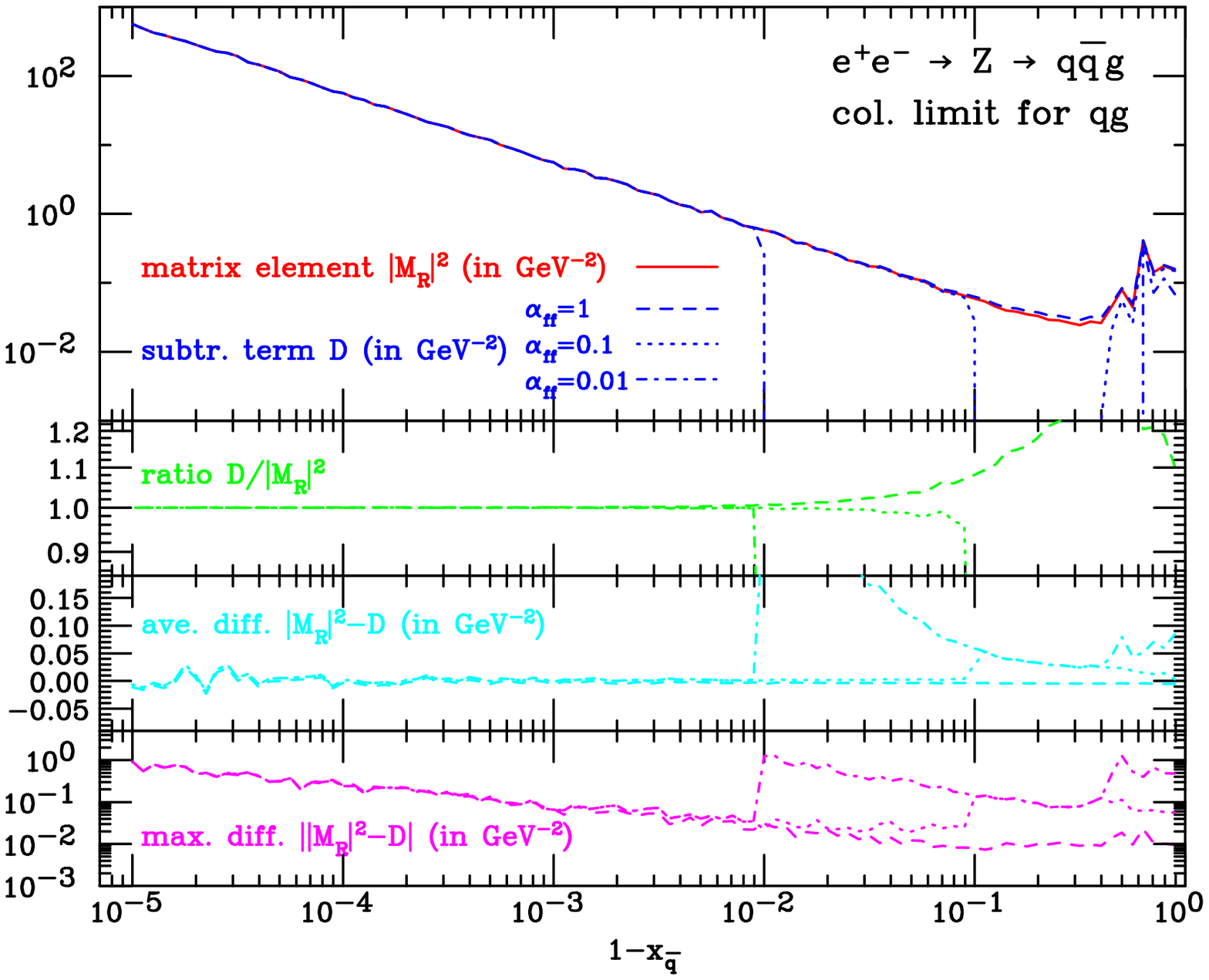, width=0.48\textwidth}
      }
      \subfigure[]{
        \epsfig{file=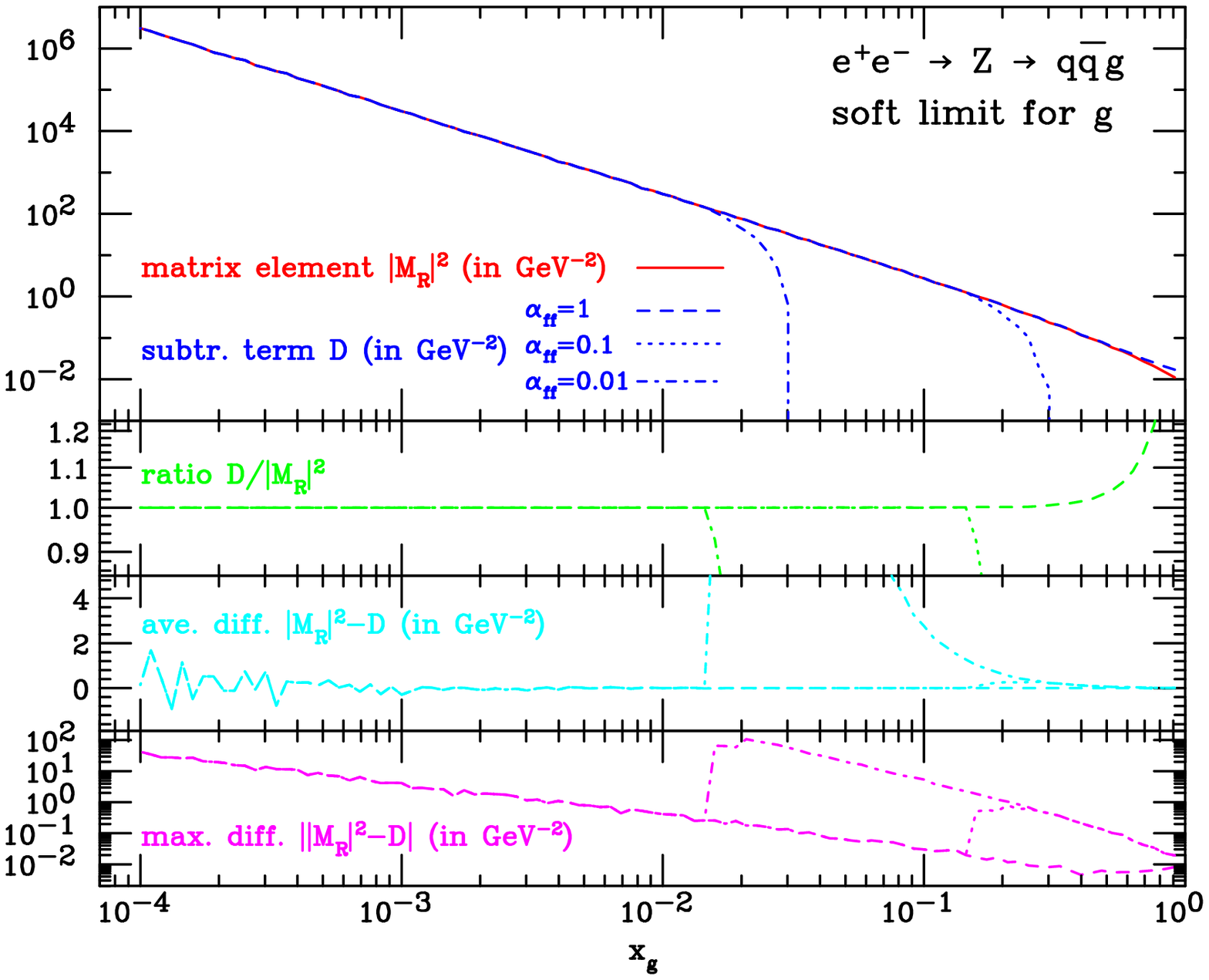, width=0.48\textwidth}
      }
    }
  \end{center}
  \vspace{-20pt}
  \caption{Matrix element squared $|M_R|^2$ (upper plots, solid line)
    and the subtraction terms $D$ (upper plots, dashed/dotted/dot-dashed lines) for the
    process $e^+(p_1)e^-(p_2)\to Z \to q(p_3)\bar{q}(p_4)g(p_5)$ as a function of
    $1-x_{\bar{q}}=(p_3.p_5)/(p_1.p_2)$ and $x_g=1-(p_3.p_4)/(p_1.p_2)$ in figures (a) and (b),
    respectively.
    Also plotted are the ratio $D/|M_R|^2$,
    the difference $|M_R|^2-D$ (averaged over 100 random points per
    bin) and the maximal difference $\textrm{max}(||M_R|^2-D|)$ per bin.
    The dashed lines include the dipoles for each point in phase space, $\alpha_{ff}=1$,
    while for the dotted $\alpha_{ff}=0.1$ and dot-dashed $\alpha_{ff}=0.01$ the
    phase space for the dipoles has been restricted to the collinear/soft regions.}
      \label{e+e-_Z_uug}
\end{figure}

\begin{figure}[t]
  \begin{center}
    \mbox{
      \subfigure[]{
        \epsfig{file=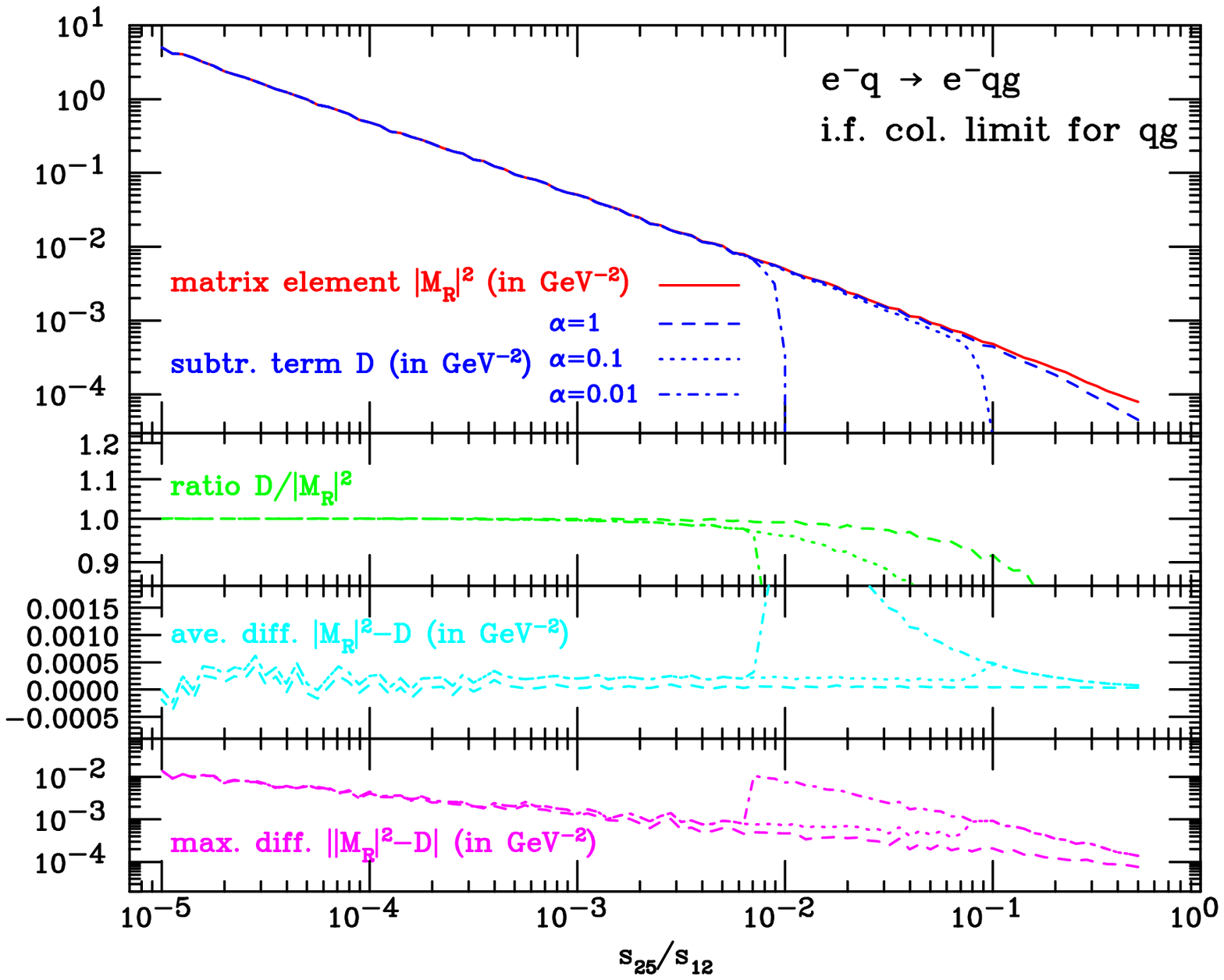, width=0.48\textwidth}
      }
      \subfigure[]{
        \epsfig{file=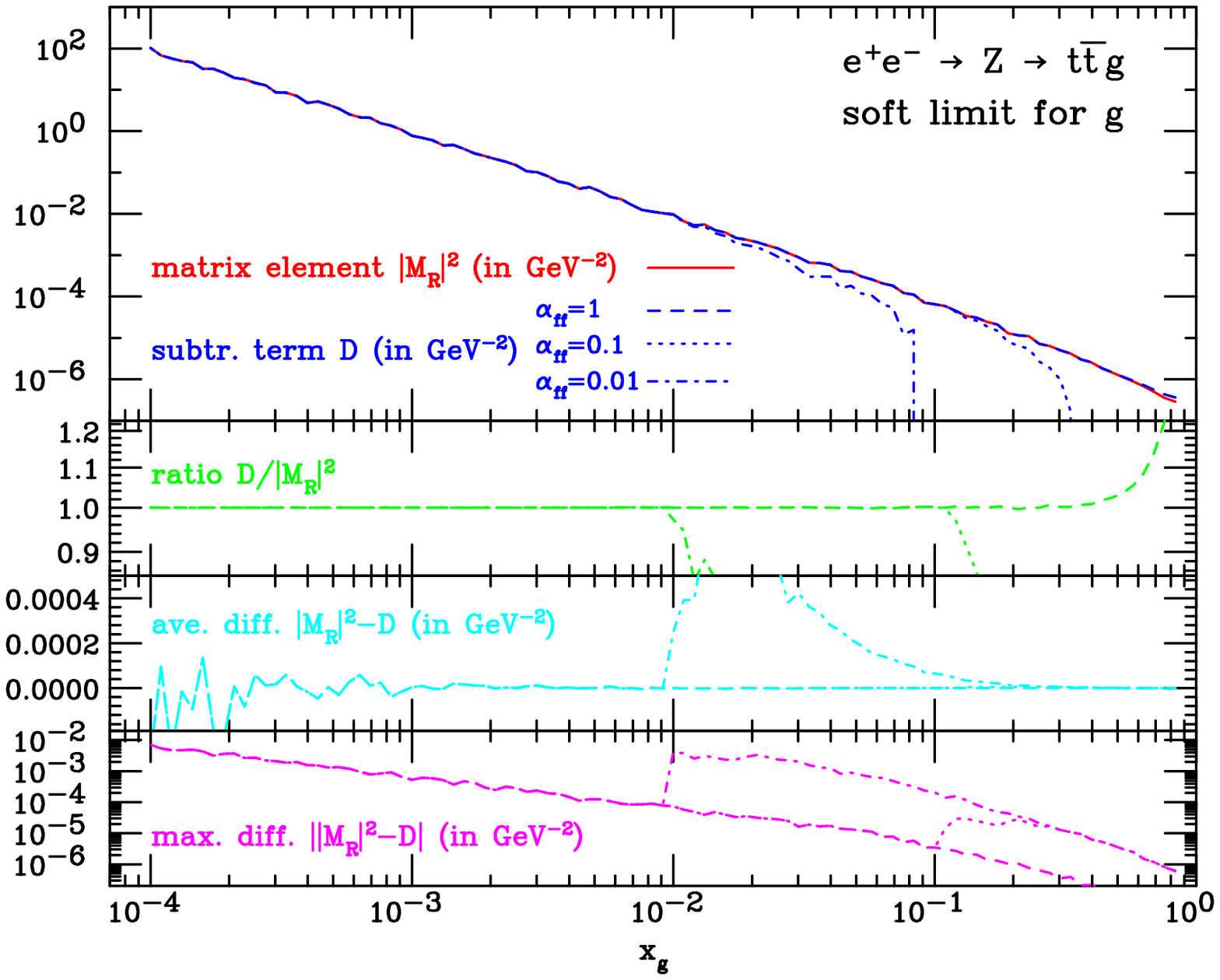, width=0.48\textwidth}
      }
    }
  \end{center}
  \vspace{-20pt}
  \caption{Matrix element squared $|M_R|^2$ (upper plots, solid line)
    and the subtraction term $D$ (upper plots,
    dashed/dotted/dot-dashed lines) for (a) the process
    $e^-(p_1)q(p_2)\to e^-(p_3)q(p_4)g(p_5)$ as a function of
    $s_{25}/s_{12}=(p_2.p_5)/(p_1.p_2)$ and (b) the process
    $e^+(p_1)e^-(p_2)\to Z \to t(p_3)\bar{t}(p_4)g(p_5)$ as a function
    of $x_{g}=1-(p_3.p_4)/(p_1.p_2)$.  Also plotted are the ratio $D/|M_R|^2$, the
    difference $|M_R|^2-D$ (averaged over 100 random points per bin)
    and the maximal difference $\textrm{max}(||M_R|^2-D|)$ per bin.
    The dashed lines include the dipoles for each point in phase
    space, $\alpha=1$, while for the dotted $\alpha=0.1$ and
    dot-dashed $\alpha=0.01$ the
    phase space for the dipoles has been restricted to the collinear/soft regions.}
      \label{init-mass}
\end{figure}

The MadDipole package provides  a 
code, \texttt{check\_dip.f}, which allows  the user to test the
limits of the $(n+1)$-particle matrix element and the dipole subtraction
terms. This code builds up a trajectory of randomly selected phase space 
point approaching a given soft or collinear limit of the $(n+1)$-parton 
matrix element and yields the values of matrix element, 
sum of all the dipoles, 
and their ratio along this trajectory. 
The result is printed to the screen in a small table for which each
successive row is closer to the singularity. The ratio between matrix
element and the sum of the dipoles should go to unity. We
have tested our code in all possible limits, both for massless as well
as massive dipoles and found no inconsistencies. Choosing small values
for $\alpha$-parameters, {\it e.g.}, $\alpha= 0.1$, improves the
computation time and the convergence of the subtraction procedure.\\
To show that the subtraction terms are implemented correctly we provide
a couple of examples in the form of plots and argue that the cancellation between
the matrix element squared and the subtraction term is as expected.
In the figures \ref{e+e-_Z_uug} and \ref{init-mass} we show the matrix
element squared and the subtraction term as a function of a variable
that represents a soft and/or collinear limit of the process
specified.  For these figures we have binned the $x$-axis (equally
sized bins for the logarithmic scale) and generated random points in
phase space to fill each of the bins with exactly 100 events.  In the
upper plot, $|M_R|^2$ and $D$ are the per bin averages of the matrix
element squared and the subtraction term, respectively. The second to
upper plot shows the per bin average of the ratio of the matrix element
squared and the subtraction term, while the third plot from the top
shows the per bin average of the difference. The lowest plot shows the
absolute value of the maximal difference among the 100 points in a
bin. To show the effects of the phase space restriction for the 
dipoles, see section \ref{PSrestriction}, all the plots
are given for $\alpha=1$ (dashed lines), $\alpha=0.1$ (dotted lines)
and $\alpha=0.01$ (dot-dashed lines).\\
In figure \ref{e+e-_Z_uug}(a) we show the matrix element
squared and the subtraction term as a function of $1-x_{\bar{q}}$,
where $x_{\bar{q}}$ is the fraction of the energy carried by the
anti-quark, $x_{\bar{q}} =\frac{s_{34}+s_{45}}{s_{12}}
=1-\frac{s_{35}}{s_{12}}$, with $s_{ij}=p_i.p_j$. For this process,
$e^+(p_1)e^-(p_2)\to Z \to q(p_3)\bar{q}(p_4)g(p_5)$, there are only final-final
state dipoles contributing to the subtraction term. The center of
mass energy is set equal to the $Z$ boson mass $\sqrt{s}=m_Z$. To
restrict the discussion to the collinear divergence only, points close
to the soft divergence ($x_{\bar{q}}=x_{\bar{q}}=1$)
have been removed by forcing $x_{q}+x_{\bar{q}}< 1.5$ in the
generation of the phase space points.\\
From the upper plot it is clear that both the matrix element squared
and the subtraction term diverge in the collinear limit
$x_{\bar{q}}\to 1$, as $1/x_{\bar{q}}$.  The ratio $D/|M_R|^2$ goes to
$1$ and the average values of the differences fluctuate close to $0$ as
can be seen in the second and third plots from the top.  The numerical
fluctuations for small $1-x_{\bar{q}}$ can be completely explained by
statistical fluctuations. They are of the order of 1\% of the maximal
difference given in the lower plot.  As can be expected, the
cancellations are not exact, which is shown by the lower plot. The
maximal difference between $|M_R|^2$ and $D$ rises like
$1/\sqrt{1-x_{\bar{q}}}$, which does not lead to a divergent phase
space integral, because the integration measure is proportional to
$x_{\bar{q}}$.
The small peaks/fluctuations in the region for small $x_{\bar{q}}$ are
due to the fact that we are 
approaching the other collinear limit, {\it i.e.}, for which
the gluon is collinear to the anti-quark $x_q\to 1$, where
the matrix elements squared and the subtraction term also
diverge.\\
In figure \ref{e+e-_Z_uug}(b) the same matrix elements and
subtraction terms are presented, but as a function of the fraction of the
energy carried away by the gluon $x_g=2-x_q-x_{\bar{q}}$. The limit
for which $x_g$ goes to zero represents the soft divergence of this
process, while the collinear divergences for this process are removed by
excluding phase space points for which $x_{\bar{q}}>(1+x_q)/2$ or
$x_{q}>(1+x_{\bar{q}})/2$. The same conclusion as for figure
\ref{e+e-_Z_uug}(a) can be drawn here: the matrix element squared and
the subtraction term diverge in the soft limit, their ratio goes to
one and the average difference to zero, while the absolute value of
the maximal difference still rises when approaching the soft limit, but does
not lead to a divergent phase space integral.\\
An example for a collinear limit between final state and initial state particles
is given in figure \ref{init-mass}(a). In this plot the matrix element squared
and the subtraction term for the process $e^-(p_1)q(p_2)\to e^-(p_3)q(p_4)g(p_5)$
are given as a function of the invariant mass of the initial state quark and
the final state gluon $s_{25}/s_{12}$.
As this invariant mass goes 
to zero, the matrix element squared and the subtraction term diverge like
$1/s_{25}$, and their ratio goes to one. To remove the other possible
divergences a cut on the momentum transfered $s_{13}/s_{12}>0.5$ and
on the invariant mass of the final state quark and gluon $s_{45}/s_{12}>0.2$ have been imposed.
Like for the final-final state dipoles
the average difference goes to a constant, as can be seen from the second plot
from the bottom, but the dipoles have a sizeable constant contribution.
Therefore the normalization of the average value for the difference $|M_R|^2-D$
depends on the number of dipoles included for the phase space point.
If all the dipoles are included for all points
the difference goes to a smaller constant than if we restrict the phase space of the subtraction term
to be close to the singularities by setting $\alpha<1$. Due to this restriction
only the dipoles to cancel that divergence are included in the 
subtraction term and therefore give a smaller constant contribution, hence the
difference $|M_R|^2-D$ is larger.
Also here the maximal left-over difference, the lowest plot, increases for
small invariant masses but does not lead to a divergent phase space integral.\\
In figure \ref{init-mass}(b) an example with massive final state particles is shown.
The process is $t\bar{t}$ production at a linear collider,
$e^+(p_1)e^-(p_2)\to t(p_3)\bar{t}(p_4)g(p_5)$ at 1 TeV center of mass energy. The
plot shows a behavior very similar to the massless case, fig.~\ref{e+e-_Z_uug}, and 
the conclusions drawn there apply to this plot as well.\\
\begin{table}
\begin{center}
\small{
\begin{tabular}{|c|c|}
\hline
process & subprocesses \\
\hline
Drell-Yan ($W$)& $q\bar{q}'\to W^+(\to e^+\nu_e)g$\\
               & $qg\to W^+(\to e^+\nu_e)q'$\\
\hline
Drell-Yan ($Z$)& $q\bar{q}\to Z(\to e^+e^-)g$\\
               & $qg\to Z(\to e^+e^-)q$\\
\hline
Drell-Yan ($Z$+jet)& $q\bar{q}\to Z(\to e^+e^-)q'\bar{q}'$\\
                   & $q\bar{q}\to Z(\to e^+e^-)q\bar{q}$\\
                   & $q\bar{q}\to Z(\to e^+e^-)gg$\\
                   & $q\bar{g}\to Z(\to e^+e^-)qg$\\
                   & $g\bar{g}\to Z(\to e^+e^-)q\bar{q}$\\
\hline
top quark pair ($t\bar{t}$)& $q\bar{q}\to t(\to bl^+\nu_l)\bar{t}(\to \bar{b}l^-\bar{\nu}_l)g$\\
                           & $qg\to t(\to bl^+\nu_l)\bar{t}(\to \bar{b}l^-\bar{\nu}_l)q$\\
                           & $gg\to t(\to bl^+\nu_l)\bar{t}(\to \bar{b}l^-\bar{\nu}_l)g$\\
\hline
$t$-channel single top & $gg\to t\bar{b}q\bar{q}'$\\
with massive $b$-quark\cite{CFMT}& $qq'\to t\bar{b}q'q''$\\
                       & $qq'\to t\bar{b}q'q''$\\
                       & $qg\to t\bar{b}q'g$\\
\hline
\end{tabular}
}
\end{center}
\caption{Set of processes for which the MadDipole code has been tested against MCFM for
random points in phase space. All the possible initial-initial, initial-final and final-initial,
dipoles for massless and massive final state particles have been
checked with this set of subprocesses.
No inconsistencies were found.}
\label{comp}
\end{table}
As a further check we have tested the code extensively against
MCFM~\cite{Campbell:1999ah,Campbell:2002tg}. 
We have generated random points in phase
space and compared the subtraction terms calculated by MCFM with the
subtraction terms calculated by our code. See table~\ref{comp} for a list
of processes that have been checked.
We observed differences only in 
the case where dipoles were introduced entirely to cancel collinear limits, 
which can be made independently of the spectator particle. 
In our code all possible dipoles are
calculated, which implies a sum over all 
spectator
particles. However, if there is only a collinear divergence, {\it i.e.},
the unresolved parton cannot go soft, this sum is
redundant and one dipole with the appropriate coefficient is enough to
cancel the singularity. In MCFM, these special limits are implemented 
using a single spectator momentum, while MadDipole sums over all spectator 
momenta, thereby yielding a different subtraction term. We have checked 
in the relevant cases that close to the singularities the MCFM subtraction 
terms behave identical to the subtraction terms calculated by our code. \\
We also tested the CPU time which is needed to produce the squared
matrix element and the dipoles for a given phase space point. These checks 
were performed with an Intel Pentium 4 processor with 3.20Ghz. As an example
we picked out three different processes:
\begin{itemize}
\item [1)] $gg\ \rightarrow \ gggg$: \ $|{\cal M}|^2$:\ 26ms, \quad $\sum_{dipoles}$:\ 68ms
\item [2)] $u\bar{u}\ \rightarrow \ d\bar{d}ggg$: \ $|{\cal M}|^2$:\ 10ms, \quad $\sum_{dipoles}$:\ 45ms
\item [3)] $u\bar{u}\ \rightarrow \ u\bar{u}ggg$: \ $|{\cal M}|^2$:\ 34ms, \quad $\sum_{dipoles}$:\ 0.15s
\end{itemize}
The time which is needed to produce the \texttt{Fortran} code is 
strongly dependent on the process and ranges from a few seconds to at most
a few minutes. The process $gg\rightarrow 5g$ is currently not yet feasible within 
MadGraph because of the size of the color factors. Once MadGraph
has been adjusted to handle this process, it will equally become 
accessible for MadDipole.

\section{How to use the package}
\label{sec:use}
The installation and running of the MadDipole package is very similar
to the usual Stand-Alone version of the MadGraph code.  In this
section, we only provide a brief description for MadDipole, while more
information can also be found on the MadGraph wiki page,
\texttt{http://cp3wks05.fynu.ucl.ac.be/twiki/bin/view/Software/MadDipole}.
\begin{itemize}
\item[1.] Download and extract the MadDipole package,
\texttt{MG\_ME\_DIP\_V4.4.3.tar.gz}, from one of the MadGraph
websites, {\it e.g.}, \texttt{http://madgraph.hep.uiuc.edu/}.
\item[2.] Run \texttt{make} in the \texttt{MadGraphII} directory.
\item[3.] Copy the \texttt{Template} directory into a new directory,
{\it e.g.}, \texttt{MyProcDir} to ensure that you always have a clean
copy of the Template directory.
\item[4.] Go to the new \texttt{MyProcDir} directory and specify your
process in the file \texttt{./Cards/proc\_card.dat}.  This is the
$(n+1)$-particle process you require the subtraction term for.
\item[5.] Running \texttt{./bin/new\_process} generates the code for
the $(n+1)$-particle matrix element and for all dipole terms. After
running this you will find a newly generated directory
\texttt{./Subprocesses/P0\_yourprocess} ({\it e.g.},
\texttt{./Subprocesses/P0\_e+e-\_uuxg}) which contains all required
files.
\item[6.] For running the check program change to this directory and
run \texttt{make} and \texttt{./check}.
\end{itemize}
The directory \texttt{./Subprocesses/P0\_yourprocess} contains all
necessary files needed for further calculations. As in the usual
MadGraph code the $(n+1)$-particle matrix element is included in the
file \texttt{matrix.f}.  For the dipoles there exist several files
called \texttt{dipol???.f} where \texttt{???} stands for a number
starting from \texttt{001}.  Each file contains only one dipole. Note
that the syntax for calling the dipoles is exactly the same as calling
the $(n+1)$-particle matrix element.  In particular, also the dipoles
have to be called with $(n+1)$ momenta rather than with only $n$
momenta. The remapping of the momenta going from the $(n+1)$-particle
phase space to a $n$-parton phase space is done within the code for
the dipoles.\\
The file \texttt{dipolsum.f} calculates the sum over all dipoles for a
given $(n+1)$-particle phase space point. It contains two subroutines
called \texttt{DIPOLSUM} and \texttt{DIPOLSUMFINITE}. As discussed in
Section~\ref{sec:masses} above, the subroutine \texttt{DIPOLSUMFINITE}
contains the dipoles which only contribute potentially large
logarithms but not a real singularity.\\
In both subroutines the contribution of a certain dipole is only added
to the sum if the phase space restriction specified by the $\alpha$
parameter is fulfilled.  The value of the four $\alpha$ parameters can
be changed in these two subroutines. They are all set to unity by
default.

\section{Conclusions}
\label{sec:conc}
In this paper we have presented MadDipole, an implementation to fully
automatize the calculation of the dipole subtraction formalism for
massless and massive partons in the MadGraph/MadEvent framework.  The
implementation is done in such a way that the user only needs to
specify the desired $(n+1)$-particle process and our code returns a
Fortran code for all dipoles combined with possible Born processes
which can lead to the $(n+1)$ process specified by the user.\\
For the calculation of the new color factors we have used as far as
possible the routines already provided by the original MadGraph code.
We inserted the two additional operators for emitter and spectator and
modified the evaluation of the squared color factors.\\
For the contributions that are not diagonal in helicity space we again
used the already available routines for calculating amplitudes for a
given helicity combination and combined amplitudes with different
helicity combinations to yield the off-diagonal helicity contributions
to the subtraction terms.\\
We have validated the code on numerous different processes with
massive and massless partons using two checking procedures.  The ratio
of ${\cal M}^2_{n+1}/\sum(dipoles)$ was confirmed to approach unity as
one approaches any soft or collinear limit. The package includes a
file \texttt{check\_dip.f} which allows the user to reproduce this
check for any process under consideration.\\
Secondly, we compared our code against the results of 
MCFM~\cite{Campbell:1999ah,Campbell:2002tg}, where 
subtraction is performed using the dipole formalism, finding point-wise 
agreement wherever anticipated. Differences with MCFM are understood to 
be due to different details in the implementation. \\
The MadDipole package allows the automated computation of  real radiation 
dipole subtraction terms required for NLO calculations. Together with 
the fastly developing automation of one-loop calculations of multi-leg
 processes, it could lead to a full automation of NLO calculations for 
collider processes. In view of the large number of potentially relevant 
multi-particle production processes at LHC, such automation will be 
crucial for precision phenomenology, in order to establish and 
interpret potential deviations from standard model expectations.

\acknowledgments{We would like to thank Fabio Maltoni and Keith Hamilton for
useful comments and discussions. NG wants to thank Gabor Somogyi for useful
discussions and comparing results. He also would like to thank the University of Louvain
for kind hospitality where part of this work was done. 
This work was  partially supported by the
Belgian Federal Science Policy (IAP 6/11)  and 
 by the Swiss National Science Foundation (SNF)
under contract 200020-117602.}

\bibliographystyle{utphys}
\bibliography{references}
\end{document}